\documentclass[superscriptaddress,twocolumn]{revtex4-1}
\usepackage{graphicx}% Include figure files
\usepackage{dcolumn}% Align table columns on decimal point
\usepackage{bm}% bold math
\usepackage[latin1]{inputenc}
\usepackage{amsmath}
\begin{document}

\title{ Investigating the `Past of a Particle' without disturbing it }
\author{F.A. Hashmi}
%\email{faheel4@comsats.edu.pk}
 \affiliation{Department of Physics, COMSATS Institute of Information Technology,  Islamabad, Pakistan.}
\author{Fu Li}%
\affiliation{Institute of Quantum Science and Engineering (IQSE) and Department
of Physics and Astronomy, Texas A\&M University, College Station, Texas, 77843-4242, USA}
\author{Shi-Yao Zhu}%
\affiliation{%
Department of Physics, Zhejiang University, Hangzhou, P.R. China}
 \date{\today}
%\keywords{Two State Vector Formalism, weak measurements, interference effects}
%\pacs{03.65.Ta,03.65.Ca,42.25.Hz}
%%%%%%%%%%%%%%%%%%%%%%%%%%%%%%%%%%%%%%% Definitions 

\newcommand{\ket}[1]{\left|#1\right\rangle}
\newcommand{\bra}[1]{\left\langle#1\right|}
\newcommand{\exptV}[1]{\left\langle#1\right\rangle}
\newcommand{\weakV}[1]{\exptV{#1}_\mathrm{w}}

\newcommand{\abs}[1]{\left|#1\right|}
\newcommand{\po}[2]{\left\langle#1|#2\right\rangle}

\newcommand{\op}[2]{\ket{#1}\bra{#2}}

\newcommand{\opR}[1]{\hat{#1}}
\newcommand{\paren}[1]{\left(#1\right)}
\newcommand{\saren}[1]{\left[#1\right]}
\newcommand{\atom}[1]{\frac{\ket{b}_{#1} + \ket{c}_{#1}}{\sqrt 2}}
\newcommand{\atomop}{\frac{e^{-i\epsilon_1}\ket{b}_{1} + \ket{c}_{1}}{\sqrt 2}}
\newcommand{\atomtp}{\frac{e^{-i\epsilon_2}\ket{b}_{2} + \ket{c}_{2}}{\sqrt 2}}

%%%%%%%%%%%%%%%%%%%%%%%%%%%%%%%%%%%%%%%%%%%%%%%%%%%
 \begin{abstract}

  In a recent article [Chin. Phys. Lett. \textbf{34}, 020301 (2017)], Ben-Israel et al.  have claimed that the experiment proposed in [Chin. Phys. Lett. \textbf{32}, 050303 (2015)] to determine the past of a quantum particle in a nested Mach-Zehnder interferometer does not work, and they have proposed a modification to the experiment. We show that their claim is false,  and the modification is not required.

\end{abstract}
\maketitle
`Past of a quantum  particle'  is a hot debate that has started with a theory \cite{Vaidman13,Vaidman14,Vaidman07} proposed by Prof. Vaidman. The theory predicts the presence of quantum particles in regions which are disconnected by the destructive interference  from the places where the particles are finally revealed. This strange prediction has been claimed in a higly debated experiment \cite{Danan13}, and the theory and the experiment have attracted considerable criticism from the community \cite{Li13,bartkiewicz_one-state_2015,nikolaev_paradox_2017,griffiths_particle_2016,potocek_which-way_2015,alonso_can_2015,svensson_non-representative_2015,saldanha_interpreting_2014,sokolovski_asking_2017,bula_measuring_2016,wu_tracing_2015,Hashmi15,Hashmi16}. The trick used in the theory \cite{Vaidman13,Vaidman14,Vaidman07,Danan13} is `the disturbance of the destructive interference' by the action of the weak measurements performed on the systems  to verify the predictions of the theory. With such disturbance the region where particles are predicted to be present in the past no longer remains disconnected from the region where it is finally detected.  The important point to note is that the prediction \cite{Vaidman13} is made for the system with  destructive interference in place and the verification \cite{Danan13} is provided for the system with disturbed destructive interference. In our previous article \cite{Hashmi15} we have proposed an experiment to investigate the past of the particle without disturbing destructive interference. Our results \cite{Hashmi15} show that the particles revealed inside the inner interferometer by the weak measurements can not continue their journey to the detector in the presence of the destructive interference on the way, and the particles post-selected on the detector do not leave any trace inside the inner interferometer. This finding is in stark contradiction to the prediction of the theory \cite{Vaidman13}. In a recent article Ben-Israel et al. \cite{Vaidman17} have criticized our experiment saying that  the experiment does not reveal the past of the particle. In the present contribution we address their criticism and show that their arguments are based on a grave misunderstanding. We also show that the authors of \cite{Vaidman17} have introduced new elements in the discussion that weaken the original claims of the theory \cite{Vaidman13}.  We first give a short introduction to our experiment \cite{Hashmi15} and then address the comments of Ben-Israel et al. \cite{Vaidman17}. 

\begin{figure}[htbp]
  \centering
  \includegraphics[width=0.5\textwidth,page=1]{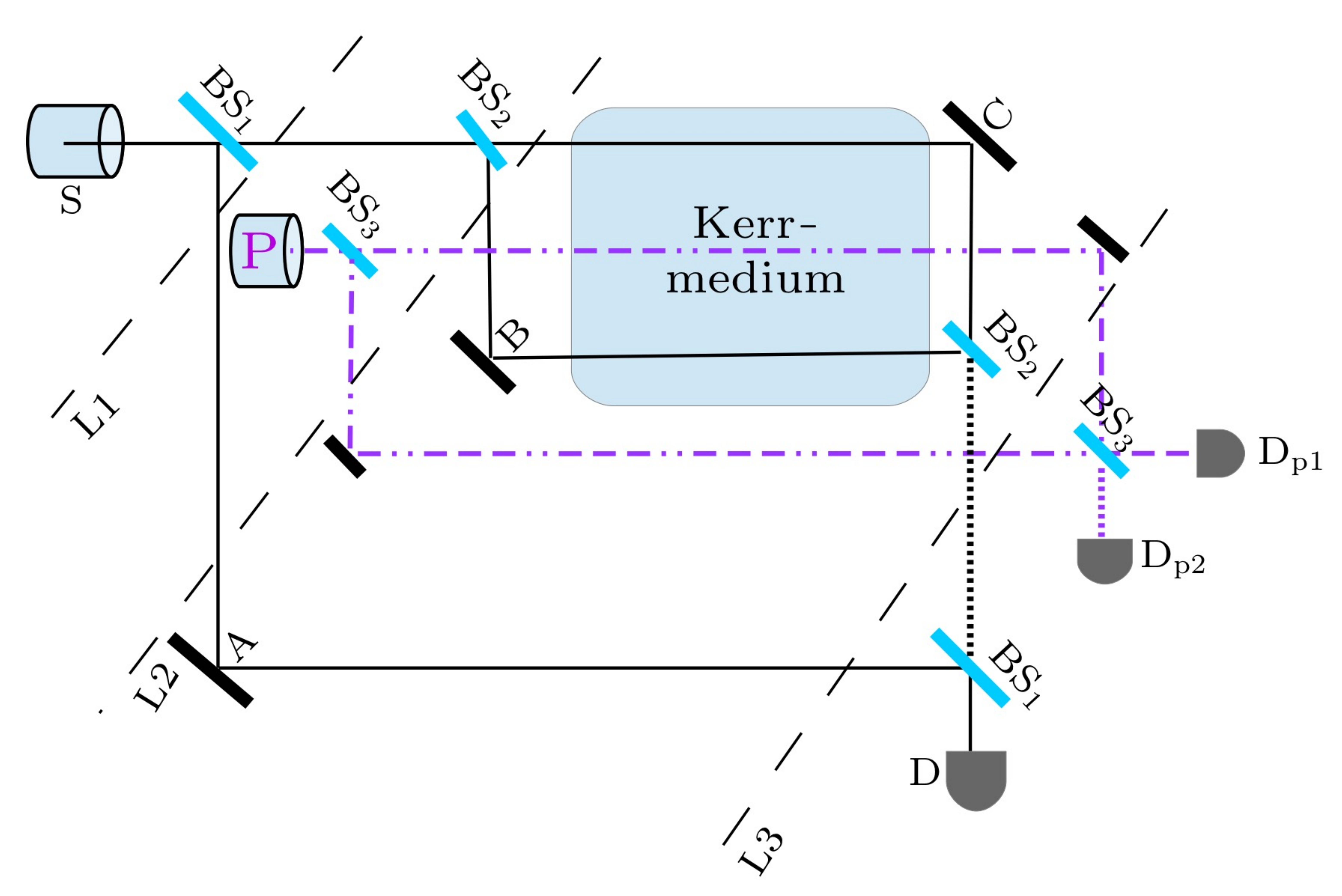}
  \caption{The nested Mach-Zehnder interferometer with a weak measurement setup taken from \cite{Hashmi15}. See text for discussion.}
  \label{fig:1}
\end{figure}

Consider the system shown in FIG.~\ref{fig:1}. The beam splitters $\textrm{BS}_1$ and the mirrors A and C constitute the outer Mach-Zehnder interferometer along one arm of which is placed an inner interferometer consisting of the beam splitters $\textrm{BS}_2$ and the mirrors B and C. The inner interferometer is set such that the path between the second $\textrm{BS}_1$ and second $\textrm{BS}_2$ (shown dotted in the figure) is a dark port. Thus a photon from inside the inner interferometer can not continue its journey towards the detector D because of this destructive interference on the way. However, the overlap of the forward and backward evolving state of two-state vector formalism \cite{Aharonov91} is present  inside the inner interferometer, and the theory \cite{Vaidman13} associates the past of a photon detected at the detector D with the inner interferometer. The verification comes from \cite{Danan13} where a weak measurement inside the inner interferometer opens up the blocked channel making it possible for the photon to reach the detector. We have proposed a different scheme for the weak measurement of the photon inside the inner interferometer. We use a probe field passing through another Mach-Zehnder interferometer constituted by the beam splitters $\textrm{BS}_3$. The probe interacts with the system photon coming from the source S inside the inner interferometer in a Kerr-medium. The part of the probe field that interacts with the photon acquires a phase that can be detected at the detectors $\textrm{D}_{pi}$ with $i=1,2$. The novelty in the scheme is that we probe the presence of the photon  inside the inner interferometer without extracting the which path information and hence without disturbing the destructive interference on the dark port (of the inner interferometer).   This is thus a more suitable experiment to test the prediction of \cite{Vaidman13} than the one performed in  \cite{Danan13}, as the destructive interference is not disturbed during the measurement process. However, this scheme has been criticised by Ben-Israel et al. \cite{Vaidman17}. We next address their comments.

\begin{enumerate}
  \item The authors start by saying that our experiment tests the presence of the photon in (the arms) B and C together.

This is a grave misunderstanding on the part of the authors, and the entire argument of the authors is based on this misunderstanding. A test for the presence of the photon in B and C together will be the one that returns a positive result only if the photon is present along both arms B and C. This clearly is not the case in our setup. A very simple analysis will tell that our scheme detects the presence of the photon if it is present either along the arm B or C. Indeed one can block the arm B or C before the entry into the Kerr-medium and the probe will still acquire the phase by the interaction with the photon along the other arm.   

This alone should be sufficient to show that the criticism of Ben-Israel et al. \cite{Vaidman17} is baseless. However we do address  their next arguments.

\item The authors next say that a negative answer to the above question does not tell if the photons were present in B or C.

  As we have already discussed before, we are not interested in asking if the photons are present along the arm B or the arm C. Our scheme answers the question if the photons are inside the inner interferometer without asking the path information in the inner interferometer. Here we wish to recall the original claim of the theory \emph{``The photon did not enter the interferometer, the photon never left the interferometer, but it was there''} \cite{Vaidman07}. Please note that the claim \cite{Vaidman13} associates the past of the particle with the overlap of the forward and backward evolving states, and hence with the inner interferometer, without making any distinction between the path B and C.

\item The author next present an analogy of the system with the three-box paradox. We believe this is irrelevant. In the three-box paradox it is not possible to look in the box B or C without resolving which box is being looked into. Moreover, looking inside a box constitutes a strong measurement which is neither allowed not related to the discussion at hand.
\item The authors say that  if we test the presence of the photon anywhere in B or C without resolving these two paths, we are certain not to find it, since it is equivalent to testing its presence in A.

We are afraid that we do not see any logic or reason in this claim.

\item The authors next say that ``if a usual (strong) measurement of an observable performed on the pre- and post-selected system yields a particular eigenvalue with certainty, a weak measurement of this observable must yield the same value''. They then continue that our experiment is such a weak measurement of the projection onto B and C together.

 We strongly disagree with it. We have already pointed out that a weak measurement onto B and C together will be the one that detects the photon only if it is present on both arms. This is not the case in our setup.

\item Next the authors bring new elements in the discussion. They say that the weak values are additive and the sum of the weak value of the presence of the photon along the arm B and C  is zero. From here it is concluded that the influence of the photon along the arm B and C cancel each other out.

  This is a new element in the theory, and it contradicts the original claim of the theory. It should be noted that weak values are independent of weak measurement strength parameter. In our setup we can slightly change the weak coupling of the probe with the arm B or C. This will disturb the destructive interference and the particle revealed inside the inner interferometer will be able to continue its journey to the detector. This is what the original theory claimed \cite{Vaidman13}. However, with the newly introduced element, the theory \cite{Vaidman13} can  no longer explain the presence of the photon inside the inner interferometer as the sum of the weak values inside the inner interferometer is still zero.
 
  Another new element without any justification or even proper explanation is the claim that ``the pre- and post-selected photon yields a superposition of the evolutions of the probe photon which can cancel each other out''. This  again raises the question that why this `superposition of the evolutions' do not cancel each other out in the absence of any weak measurements inside the inner interferometer for which the original prediction has been made.

\item The authors also say that in standard quantum mechanics there is no concept of particle path or past of the particle. It is generally true, but the continuity and connectivity of the wavefunction through different regions between the pre- and post-selection is a very basic requirement of standard quantum theory. Letting go of this requirement is very likely to give rise  to paradoxes. The theory \cite{Vaidman13} is one such example.
 \end{enumerate}
In the remaining part of their article, the authors suggest that `an improvement' of the experiment is to make the path of the probe field closer to any one arm of the inner interferometer.  We do not agree with this suggestion as it will again disturb the destructive interference on the path coming out of the inner interferometer. However, a modification  in which the probe field interacts weakly with all arms of the outer interferometer (the one containing the inner interferometer) can be an improvement to test the claim of the theory that the particle was inside the inner interferometer but not on the paths leading to or coming out of the inner interferometer.

In conclusion we have shown that the criticism of Ben-Israel et al.  on our article \cite{Hashmi15} is based on misunderstanding, and the arguments presented are flawed. Indeed the theory of the past of the particle proposed by Prof. Vaidman is  incorrect, and our scheme \cite{Hashmi15} shows it very clearly.

%\newpage
%\bibliographystyle{apsrev}
%\bibliography{manuscript}
\end{document}